# Expressive haptics for enhanced usability of mobile interfaces in situations of impairments

**Tigmanshu Bhatnagar**
University College London Interaction Centre & GDI Hub, 66-72 Gower Street, London, UK
t.bhatnagar.18@ucl.ac.uk

**Youngjun Cho**
University College London Interaction Centre & GDI Hub, 66-72 Gower Street, London, UK
youngjun.cho@ucl.ac.uk

**Nicolai Marquardt**
University College London Interaction Centre, 66-72 Gower Street, London, UK
n.marquardt@ucl.ac.uk

**Catherine Holloway**
University College London Interaction Centre & GDI Hub, 66-72 Gower Street, London, UK
c.holloway@ucl.ac.uk

## ABSTRACT
Designing for situational awareness could lead to better solutions for disabled people, likewise, exploring the needs of disabled people could lead to innovations that can address situational impairments. This in turn can create non-stigmatising assistive technology for disabled people from which eventually everyone could benefit. In this paper, we investigate the potential for advanced haptics to compliment the graphical user interface of mobile devices, thereby enhancing user experiences of all people in some situations (e.g. sunlight interfering with interaction) and visually impaired people. We explore technical solutions to this problem space and demonstrate our justification for a focus on the creation of kinaesthetic force feedback. We propose initial design concepts and studies, with a view to co-create delightful and expressive haptic interactions with potential users motivated by scenarios of situational and permanent impairments.

## INTRODUCTION
Mobile interaction has become pervasive and leads to behaviours wherein situational impairments are commonly encountered. We have often seen people traversing a dense crowd while using their mobile device or people grabbing a coffee in one hand and texting with the other while balancing their body in the jerky motion of a public tube. Designing for situational awareness could lead to better solutions for disabled people, likewise, designing for the needs of disabled people could lead to innovations that can address situational impairments. This in turn can create non-stigmatising assistive technology for disabled people that eventually everyone could benefit from.

Mobile devices are held in one of the most sensitive cutaneous areas of the body with tactile discrimination extending to the nanoscale [15]. The constellation of mechanoreceptors, nerve endings and the primary endings of the muscle spindles allows for differentiation between a range of textures, vibrations, materials, temperatures and pressures [12]. Motivated by scenarios of situational and permanent impairments, the use of haptics to provide distinguishable input and feedback has the potential to enhance mobile HCI. In this paper, we discuss the first steps towards the design of a novel haptic force feedback mechanism and explore use cases for both situational and permanent disability.

**RELEVANT WORK - HAPTICS**

Haptic research for disabled people finds its rich and innovative niche in the domain of navigation and way finding applications for visually impaired people [2, 13]. Within this, a vast majority of experiments utilises the vibrotactile approach for creating the haptic cues for orientation and navigation, through both hand held [11] and wearable artefacts [8]. Similarly, haptics for the deaf community for proprioceptive orientation, recreating audio cues and local awareness have been developed [14]. Few studies utilise a kinaesthetic approach for a similar purpose in hand held artefacts [1, 7]. Vibrations are also utilised to render digital graphical information for visually impaired people. These surface haptics, utilise electrovibration [3], ultrasonic vibrations [10], pen based vibrations [5] and shape changing liquid crystal elastomers [17] to generate a variety of surface texture information. However, these technologies are in their nascent years. We believe that by encompassing a wider number of use cases, inspired by situational impairment, these solutions can find a home in product ranges beyond prototype designs.

**THE CHALLENGE**

In such specialist innovations of haptic media for people with disabilities, there is a danger that the innovations can miss the opportunity to create inclusive solutions which does not stigmatise or differentiate its users [6]. Furthermore, the most widely used approach in haptic feedback is vibrotactile feedback because it is fundamentally straightforward to engineer vibrations and humans are well adapted to perceive them [4]. It is meaningful to use it to attract attention and notify, however conveying monotonous vibrations often makes a user to feel unpleasant or bored. When it comes to mobile's aesthetical user experience and navigation of digital content, which is also pervasive in the disabled community, haptic effects are under-utilised [9]. For instance, the only feedback visually impaired users receive upon their actions on a mobile device is through audio screen reading and clicking sounds. A multimodal approach using haptic force feedback complimenting the audio-visual media becomes a compelling idea when a wider range of use cases are incorporated. For instance, increasing the haptic forces as the menu is scrolled followed by a gentle tug when its end is reached or orienting the user's attention to a particular content on the screen by manipulating the source of haptic feedback, thereby creating a delightful and purposeful interaction mode for all.

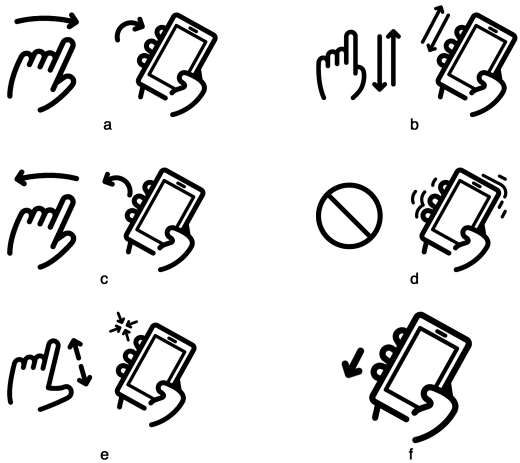

Figure 1: Various haptic expressions that can compliment the graphical and voice feedback based on gestural inputs [17]. In a, b and c the phone responds by giving a force on the hand grasping the device. This can be particularly useful for eyes off interaction while navigating with the device and for visually impaired users with their gestural inputs. d. shows that the device can also haptically interrupt without the need for a visual cue and notify the user to put their attention elsewhere. In e, the haptic feedback compliments the graphical feedback by locating the point of interest on the screen and in f. during single hand operation of the device, the CoG shifts near the palm of the hands, providing a more stable grip.

**INITIAL CONCEPT**

The initial concept aims to create 'Haptic Expressions' by a force generated through changing centre of gravity (CoG) of the mobile device in synchronisation with a variety of swiping gestures on the touch screen (Figure 1). The response can provide an informed reaction to the action performed by the user, thereby enhancing the use of mobile devices. For example, in digital navigation reaching the end of a menu is indicated by a 'visual tug' of the screen. The haptic feedback can complement this tug with a co-relatable physical force, which will indicate the similar information. Repetitive forces in a certain frequency in a certain direction can help in providing cues for real world navigation that are especially useful in scenarios with 'eyes-off' interaction during walking navigation and visually impaired navigation. The frequency of the change of CoG can correlate to the number and frequency of steps or in some cases to breathing or to the heartbeat, creating a humanistic feedback. Changing CoG can also very subtly assist in on screen navigation and provide a haptic feedback to gestures such as pinching to a point of interest, with the haptics moving to that point in sync. For single hand mobile operation, the lowering the centre of gravity of the device can create a more stable grip, which can also be helpful for the deaf community to speak one hand sign language.

The design challenge is to generate sufficient forces to provide a perceivable kinaesthetic feedback in a slim form factor and small weight. We plan to achieve this by shifting a thin layer of dense mass in the plane of the mobile device at a rapidly changing acceleration. At first, we aim to build and experiment with 2 degree of freedom manipulation mechanisms to achieve the desired weight change in a minimal form factor. These could be achieved through parallel manipulator robotic arms, an X-Y stage, piezoelectric ultrasonic motors under the mass or solenoid based actuation.

The primary research objective is to investigate human haptic perception of change of CoG to optimize the parameters for embodiment design and the accuracy of feedback. Studies are being designed but the parameters we are interested in are:
1. The weight of the shifting mass with respect to the overall weight of the device,
2. The minimum distance to be moved for perceivable change and its correlation to the weight
3. Its accelerations to generate a clearly distinguishable force
4. Patterns of movement to complement the gestures in various scenarios
The outcomes from the investigation of such parameters will contribute to designing our compact prototype, a slim 'haptic case' which could embody the technology and be added on to mainstream mobile devices. Through the medium of the prototype case, our secondary research objective is to identify and design for the scenarios where this understanding can be useful. For this, we would co-create feedback instantiations with potential users and disabled people. The session(s) is expected to lead to ideas to compliment the mobile interaction in real life scenarios. The ideas will then be designed, embodied and evaluated.

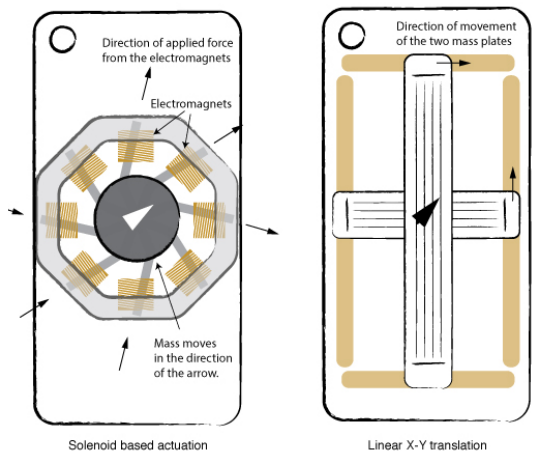

Figure 2: Representation of some of the proposed conceptual mechanisms to achieve weight shifting in order to produce a haptic feedback. The Solenoid based actuation, multiple solenoids arranged in an octagonal shape couple their forces upon excitation in order to move a mass (round grey circle) at high acceleration thereby creating a change in the CoG and a haptic reaction force. In the second concept, a similar effect is created by moving two orthogonal masses on a channel.

**CONCLUSION**

The potential for haptic force feedback in mobile interactions for disabled users is under-utilised. Its utilisation in scenarios of situational impairments and can lead to the development of non stigmatizing and inclusive assistive technology. Our initial analysis led us to experiment with generating force feedback by shifting the centre of gravity of the mobile device in a slim form factor of a haptic case. It is expected that in synchronisation with the graphical and audio interfaces, this haptic modality will enhance user's gestural interactivity, especially of visually impaired people. This technology has the potential to aid all mobile phone users, including people with disabilities.

**TOPICS FOR DISCUSSION IN THE WORKSHOP**

With the technical components this paper addresses, we would like to intensely discuss the following topics within a group or between groups in this workshop:

- Are there overlapping effects of permanent and situational disabilities? Can addressing one lead to design assistive technology that eventually everyone uses and benefits from, thereby designing for disabled people in a non-stigmatising way?
- Can there be a design method(s) through which we can invent technology that can help adapting to technology induced impairments? How might we evaluate those inventions so that they do not create new impairments?
- Is there an opportunity for multimodal interactivity? In situations of impairments, how does the engagement of multiple sensory inputs affect user experience?

**REFERENCES**

[1] Tomohiro Amemiya and Hiroaki Gomi. 2014. Distinct Pseudo-Attraction Force Sensation by a Thumb-Sized Vibrator that Oscillates Asymmetrically. Haptics: Neuroscience, Devices, Modeling, and Applications Lecture Notes in Computer Science (2014), 88–95. DOI:http://dx.doi.org/10.1007/978-3-662-44196-1_12

[2] Shiri Azenkot, Richard E. Ladner, and Jacob O. Wobbrock. 2011. Smartphone haptic feedback for nonvisual wayfinding. The proceedings of the 13th international ACM SIGACCESS ASSETS 11 (2011). DOI:http://dx.doi.org/10.1145/2049536.2049607

[3] Olivier Bau, Ivan Poupyrev, Ali Israr, and Chris Harrison. 2010. TeslaTouch. Proceedings of the 23nd annual ACM symposium on User interface software and technology - UIST 10 (2010). DOI:http://dx.doi.org/10.1145/1866029.1866074

[4] A.J. Brisben, S.S. Hsiao, and K.O. Johnson. 1999. Detection of Vibration Transmitted Through an Object Grasped in the Hand. Journal of Neurophysiology 81, 4 (1999), 1548–1558. DOI:http://dx.doi.org/10.1152/jn.1999.81.4.1548

[5] Youngjun Cho, Andrea Bianchi, Nicolai Marquardt, and Nadia Bianchi-Berthouze. 2016. RealPen. Proceedings of the 29th Annual Symposium on User Interface Software and Technology - UIST 16 (2016). DOI:http://dx.doi.org/10.1145/2984511.2984550.

[6] Gerard Goggin. 2017. Disability and haptic mobile media. New Media & Society 19, 10 (2017), 1563–1580. DOI:http://dx.doi.org/10.1177/1461444817717512

[7] Fabian Hemmert, Susann Hamann, Matthias Löwe, Josefine Zeipelt, and Gesche Joost. 2010. Weight-shifting mobiles. Proceedings of the 28th of the international conference extended abstracts on Human factors in computing systems - CHI EA 10 (2010). DOI:http://dx.doi.org/10.1145/1753846.1753922.

[8] Wilko Heuten, Niels Henze, Susanne Boll, and Martin Pielot. 2008. Tactile wayfinder. Proceedings of the 5th Nordic conference on Human-computer interaction building bridges - NordiCHI 08 (2008). DOI:http://dx.doi.org/10.1145/1463160.1463179

[9] Vincent Levesque 2005. Blindness, technology and haptics. Center for Intelligent Machines (2005), 19-21.

[10] David J. Meyer, Michael Wiertlewski, Michael A. Peshkin, and J.Edward Colgate. 2014. Dynamics of ultrasonic and electrostatic friction modulation for rendering texture on haptic surfaces. 2014 IEEE (HAPTICS) (2014). DOI:http://dx.doi.org/10.1109/haptics.2014.6775434

[11] Martin Pielot, Benjamin Poppinga, Wilko Heuten, and Susanne Boll. 2011. A Tactile Compass for Eyes-Free Pedestrian Navigation. Human-Computer Interaction – INTERACT 2011 Lecture Notes in Computer Science (2011), 640–656. DOI:http://dx.doi.org/10.1007/978-3-642-23771-3_47

[12] D. Purves, G.J. Augustine, and D. Fitzpatrick. 2001. Mechanoreceptors specialized to receive tactile information. Neuroscience. (2001).

[13] Jun Rekimoto. 2013. Traxion. Proceedings of the 26th annual ACM symposium on User interface software and technology - UIST 13 (2013). DOI:http://dx.doi.org/10.1145/2501988.2502044

[14] Peter B. Shull and Dana D. Damian. 2015. Haptic wearables as sensory replacement, sensory augmentation and trainer – a review. Journal of NeuroEngineering and Rehabilitation 12, 1 (2015). DOI:http://dx.doi.org/10.1186/s12984-015-0055-z

[15] Lisa Skedung, Martin Arvidsson, Jun Young Chung, Christopher M. Stafford, Birgitta Berglund, and Mark W. Rutland. 2013. Feeling Small: Exploring the Tactile Perception Limits. Scientific Reports 3, 1 (2013). DOI:http://dx.doi.org/10.1038/srep02617

[16] Web Icon Set from the Noun Project

[17] N. Torras et al. 2014. Tactile device based on opto-mechanical actuation of liquid crystal elastomers. Sensors and Actuators A: Physical 208 (2014), 104–112. DOI:http://dx.doi.org/10.1016/j.sna.2014.01.012